\documentstyle[12pt]{article}
\def\aprle{\buildrel < \over {_{\sim}}}   

\begin{document}   
\thispagestyle{empty}
\begin{flushright}
{\tt hep-ph/0210071}\\
{ROMA1-1342/2002}
\end{flushright}
\vspace*{1cm}
\begin{center}
{\Large{\bf A comment on hadronic charm decays} }\\
\vspace{.5cm}
M. Lusignoli \footnote{maurizio.lusignoli@roma1.infn.it}
and A. Pugliese \footnote{alessandra.pugliese@roma1.infn.it}
 
\vspace*{0.5cm}
 Dipartimento di Fisica, Universit\`a di Roma ``La Sapienza'',\\
and I.N.F.N., Sezione di Roma,\\
P.le A. Moro 2, I-00185, Rome, Italy

\end{center}

\begin{abstract}
    We give arguments in favor of the compatibility with standard 
    physics of some large nonleptonic branching fractions in 
    Cabibbo--forbidden $D^+$ decays, contrary to a recent claim in 
    the literature.
\end{abstract}

\vspace{0.2cm}

More than twenty years ago, the large difference between the lifetimes 
of charged and neutral $D$ mesons has been attributed to the effect of 
the interference between color-favored and color-suppressed 
amplitudes in $D^+$ decays \cite{Guberina:1979xw} and/or of the $W$-exchange 
contribution in the $D^0$ decay amplitudes \cite{Bander:jx}. Since the 
second explanation would imply (in its simplest form) a large 
annihilation contribution in $D_s^+$ and large branching ratios 
(BR's) for unobserved decay channels, the interference is commonly 
believed to be the main effect \cite{Shifman:1987nd}. 
Concerning exclusive decays, the surprisingly large BR for the decay 
$D^0 \to {\overline K}^0\;\pi^0$ showed that the colour suppression 
is much less effective than expected \cite{Cabibbo:1977zv,Fakirov:ta} 
in these decays. 
This fact may further point to important cancellations due to interference 
between color-allowed and color-(not-so-much)suppressed 
amplitudes in $D^+$ decays. 

All the above applies to the (dominant) Cabibbo--allowed 
decays. In the Cabibbo--forbidden $D^+$ decays to two 
strange mesons, only the color-favored tree amplitude contributes (as 
it is also the case for $D^0$ decays in two charged particles).  
Therefore, the branching fractions for these decays might be 
considerably larger than what would be obtained with the simple--minded 
suppression by a factor $\tan^2 \theta_C$ with respect to 
Cabibbo--allowed decay rates. 
The same is true for the doubly--forbidden decays $D^+ \to K^{(*)+}\;X$. 
The precise values of these 
BR's will however be affected by other effects, such as annihilation 
contributions and 
final state interactions (rescattering), and are therefore very 
model-dependent.    

In a recent letter, Close and Lipkin \cite{Close:2002fa} suggested that the 
measured BR's of the decays $D^+ \to K^{*+}{\overline K}^0$ 
($3.1 \pm 1.4\;\%$) \cite{Frabetti:1994wu} 
and $D^+ \to K^{*+}{\overline K}^{*0}$ ($2.6 \pm 1.1\;\%$) \cite{Albrecht:1991zb}, 
if confirmed with smaller error bars, may require ``new physics'' to explain them. 
We believe that this conclusion is not necessary. 

In ref.\cite{Close:2002fa} two points have been made: (a) the 
Cabibbo--forbidden branching ratios for $D^0$ decays are $\aprle
0.4\%$, namely considerably smaller than the above--mentioned BR's; (b) the 
Cabibbo--allowed $D^+$ BR's are instead not much larger. 
On the first point, we note that the ratio of lifetimes is 
$\tau_{D^+} / \tau_{D^0} = 2.55 \pm 0.04$: therefore, for equal 
decay amplitudes, a $D^+$ Cabibbo--forbidden BR of $\sim 1\%$ may be  
actually expected. 

The second point is less clear--cut, due to the effect of 
interference in the decay amplitudes. Since interferences suppress both 
numerator and denominator in the BR in this case, a simplistic idea would 
attribute similar values for the BR of $D^+$ and $D^0$ in 
Cabibbo--allowed decays: this is at variance with the experimental 
data, since BR($D^0 \to \rho^+ K^-) = 10.2 \pm 0.8\;\%$ and
BR($D^+ \to \rho^+ {\overline K}^0) = 6.6 \pm 2.5\;\%$ 
\footnote{It may be noted that $6.6 \pm 2.5\;\%$ is compatible, 
within two standard deviations, both with $3.1 \pm 1.4\;\%$ ($D^+$ 
Cabibbo--suppressed, a surprise) and with $10.2 \pm 0.8\;\%$ ($D^0$ 
Cabibbo--allowed, as roughly expected).}. 
Another embarassing fact is the considerable difference in the BR's 
for $D^+ \to K^+ {\overline K}^{*0}$ and $D^+ \to K^{*+}{\overline K}^0$.
Here hadronization effects are different, since in one case the 
virtual $W^+$ gives rise to a pseudoscalar meson, in 
the other to a vector meson. In particular models, the 
data may (or may not) be reproduced.

Models that are able to describe the main features of the two-body $D$ decay 
data without introducing any ``new physics'' have appeared 
in the literature 
\cite{Bauer:1986bm,Chau:tk,Buccella:1994nf,Buccella:1996uy} 
since a long time. They are generally 
based on a modified factorization approximation, in which the color 
suppression factor and annihilation and $W$-exchange contributions are 
determined by fitting some parameters to the experimental data. 
In the following, to give an 
example of a model where the BR for $D^+ \to K^{*+}\;\bar{K^0}$
turns out to be large (1.52\%), we present results from \cite{Bucc_BABAR}. 
These were obtained making a fit with 15 parameters to 
all BR data for $D$ decays in two pseudoscalar mesons (PP), a pseudoscalar and 
a vector meson (PV) and a pseudoscalar and a light scalar meson (PS) 
(corresponding to 56 experimental 
data points or upper bounds). Here, we will only 
discuss the fitted theoretical results for some particular decays, in order to 
illustrate the previous arguments. In particular, we will give results 
for decays to ${\overline K}^0$ and $K^0$ mesons separately, although in the fit 
decays in $K_S$ were considered, since the neutral kaons are 
experimentally detected through their two pion decays. 

The annihilation/$W$-exchange contribution in the factorized 
approximation ({\it i.e.} neglecting possible gluon emissions from the 
initial quarks) depends on the matrix element of the divergence of a vector 
(axial-vector) current for PP (PV) decays. This is transformed, using 
the equations of motion, to the
matrix element of a scalar (pseudoscalar) density multiplied by the 
difference (sum) of quark masses. 
As a consequence, the annihilation contributions to Cabibbo--allowed 
$D_s^+$ and to Cabibbo--forbidden $D^+$ decays are quite small, being 
suppressed by light quark masses. Relevant annihilation contributions 
are present instead for Cabibbo--forbidden $D_s^+$ 
and doubly--Cabibbo--forbidden $D^+$ decays. 

The pattern arising for the factorized amplitudes may be strongly mixed up by 
final state interactions. In this model, all the states belonging to 
the same representation of the flavor SU(3) symmetry mix among 
themselves. The rescattering is dominated by the nearby resonances, 
whose experimental masses and widths (when known) determine the 
phase-shifts\footnote
{A similar treatment of final state interactions has been 
recently presented in \cite{Cheng:2000fd}. However, in that paper the phase 
shifts have been erroneously multiplied by a factor of two.}.

The results of the fit for some $D^+$ decay channels of interest are 
presented in Table~1. In the first column we list the branching ratios that 
would result prior to rescattering corrections (${\rm BR_{NR}}$) -- not a fit 
to the data. The second column contains the final, rescattered results 
(${\rm BR_{RR}}$). 
In the third column we write the present experimental data \cite{PDG2002} that 
in some cases differ from the data that were fitted.

\begin{table}[ht]
\caption{
Branching ratios (in \%) for some nonleptonic
decays of $D^+$ mesons.}
\vspace{0.3 cm}
\begin {tabular} {| c l c c c | c l c c c |}
\hline
 & $D^+ \to$ & ${\rm BR_{NR}}$ & ${\rm BR_{RR}}$ &  ${\rm BR_{exp}}$ &  
& $D^+  \to$ & ${\rm BR_{NR}}$ & ${\rm BR_{RR}}$ & ${\rm BR_{exp}}$ \\
\hline
 & $\pi^+ {\overline K}^0$ & 2.939 & 2.939 &  --- & 
& $\rho^+ {\overline K}^0$ & 14.076 & 12.198 & --- \\
 & ~ & ~ & ~ & ~  & 
& $\pi^+ {\overline K}^{*0}$ & 0.076 & 1.996 & 1.92$\pm$0.19 \\
 & $\pi^+ \pi^0$ & 0.185 & 0.185 & 0.25$\pm$0.07  & 
& $\rho^+ \pi^0$ & 0.775 & 0.451 & --- \\
 & ~ & ~ & ~ & ~  & 
& $\pi^+ \rho^0$ & 0.012  &  0.104  & 0.104$\pm$0.018 \\
 & $K^+ {\overline K}^0$ & 1.486 &   0.764  & 0.58$\pm$0.06  & 
& $K^{*+} {\overline K}^0$ &  1.364  &   1.515  & 3.1$\pm$1.4 \\
 & ~ & ~ & ~ & ~  & 
& $K^+{\overline K}^{*0}$ & 0.653  &  0.436  & 0.42$\pm$0.05 \\
 & ~ & ~ & ~ & ~  & 
& $\ \phi\ \pi^+$ & 0.505 & 0.619 & 0.61$\pm$0.06 \\
 & $K^+ \pi^0$ & 0.083  & 0.055 & --- & 
& $K^{*+} \pi^0$ & 0.096  & 0.057  & --- \\
 & ~ & ~ & ~ & ~  & 
& $K^+ \rho^0$ & 0.012 & 0.029 & 0.025$\pm$0.012 \\
 & $K^{0} \pi^+$ &  0.057  & 0.053 & --- & 
& $K^{*0} \pi^+$ & 0.062 & 0.027  & 0.036$\pm$0.016 \\
 & ~ & ~ & ~ & ~ &  
& $K^0 \rho^+$ & 0.001  & 0.042 & --- \\
 & ~ & ~ & ~ & ~ & ~ & ~ & ~ & ~  & \\
 & $\pi^+ K_S$ & 1.088 & 1.347 & 1.38$\pm$0.09  & 
& $\rho^+ K_S$ & 6.946  & 5.820  & 3.30$\pm$1.25 \\
\hline
\end{tabular}
\end{table}

The left part of Table~1 refers to PP decays. We note that the two 
Cabibbo--forbidden decays have very different ${\rm BR_{NR}}$'s: this 
is due to the interference between ``tree'' and ``color--suppressed'' 
amplitudes, present in the $\pi^+ \pi^0$ case (as well as in the 
Cabibbo--allowed decay to $\pi^+ {\overline K}^0$) and absent for the 
$K^+ {\overline K}^0$ decay, which has therefore a much larger BR. 
A similar argument also applies to the doubly--forbidden decays, that 
do not have interference and have instead possibly large annihilation 
contributions: their BR's are of the order of $\tan^2 \theta_C \;
{\rm BR_{NR}}(D^+ \to K^+ {\overline K}^0$) and much larger than 
$\tan^4 \theta_C \; {\rm BR_{NR}}(D^+ \to \pi^+ {\overline K}^0$).
The rescattering effects are quite big, and approximately half of the 
biggest Cabibbo--forbidden BR is redistributed to the other 
channels, namely $\pi\eta$ and $\pi\eta'$.

In the right part of Table~1, it can be noted that the naturally 
paired channels (V$_i$P$_j$ and P$_i$V$_j$) have very different BR's. 
The simplest case to analize is the $K^{*+} {\overline K}^0$ and 
$K^+{\overline K}^{*0}$ pair, with no interference and a very small 
annihilation contribution. In the factorized approximation,
the first decay is induced by the $VV$ part of the 
effective hamiltonian, the second by the $AA$ part. The parameters 
entering in the calculation favor the first contribution, the main 
reason being the ratio of decay constants $f_{K^*}\;/ \;f_K \simeq 
1.4$; final state interactions further enhance this effect. 
When also interference enters the game, the smaller BR's may be 
almost completely cancelled, as it happens for the Cabibbo--allowed 
case. The rescattering effects are not very big for 
the channels having large ${\rm BR_{NR}}$, while they obviously change 
a lot the smallest ones. The general pattern is however rather 
stable. 

The results of the fit for some $D^+_s$ decay channels of interest are 
presented in Table~2.  Again, our comments refer to the results prior 
to rescattering corrections, listed in the first and fourth columns. 
It is to be noted that the branching ratios in final states containing 
a hypothetical $s\bar{s}$ pseudoscalar meson with the $\eta$ mass 
would be approximately twice the values reported in the first 
row in Table~2, with the value 
for the $\eta-\eta'$ mixing angle that we use \cite{dyak}. 

\begin{table}[ht]
\caption{
Branching ratios (in \%) for some nonleptonic
decays of $D^+_s$ mesons.}
\vspace{0.3 cm}
\begin {tabular} {| c l c c c | c l c c c |}
\hline
 & $D^+_s \to$ & ${\rm BR_{NR}}$ & ${\rm BR_{RR}}$ &  ${\rm BR_{exp}}$  & 
& $D^+_s  \to$ & ${\rm BR_{NR}}$ & ${\rm BR_{RR}}$ & ${\rm BR_{exp}}$ \\
\hline
 & $\pi^+  \eta$ & 5.026 & 1.131 &  1.7$\pm$0.5 & 
& $\rho^+ \eta$ & 9.182 & 8.122 & 10.3$\pm$3.2 \\
 & ~ & ~ & ~ & ~  & 
& $\pi^+ \phi$ & 5.065 & 4.552 & 3.6$\pm$0.9 \\
 & ${\overline K}^0 K^+$ & 3.285 & 4.623 & ---  & 
& ${\overline K}^{*0} K^+$ & 4.108 & 4.812 & 3.3$\pm$0.9 \\
 & ~ & ~ & ~ & ~ &  
& ${\overline K}^0 K^{*+}$ & 1.770  &  2.467  & --- \\
 & $\pi^+ K^0$ & 1.099 &   0.373  & $< 0.8$ & 
& $\rho^+ K^0$ &  0.847  &   1.288  & --- \\
 & ~ & ~ & ~ & ~  & 
& $\pi^+ K^{*0}$ & 0.898  &  0.445  & 0.65$\pm$0.28 \\
 & $\pi^0 K^+$ & 0.211  & 0.146 & --- & 
& $\rho^0 K^+$ & 0.018  & 0.198  & $< 0.29$ \\
 & ~ & ~ & ~ & ~  & 
& $\omega K^+$ & 0.288 & 0.178 & --- \\
 & ~ & ~ & ~ & ~  & 
& $\pi^0 K^{*+}$ & 0.177 & 0.076 & --- \\
 & $K^+ \eta$ &  0.003  & 0.300 & --- & 
& $K^{*+} \eta$ & 0.288 & 0.146 & --- \\
 & ~ & ~ & ~ & ~ & 
& $K^+ \phi$ & 0.014 & 0.008  & $< 0.05$\\
 & $K^+ K^0$  &  0.012 &  0.012 & --- & 
& $K^{*+} K^0$ & 0.021 &  0.018 &  --- \\
 & ~ & ~ & ~ & ~ &  
& $K^+ K^{*0}$ & 0.003 &  0.006 & --- \\
 & ~ & ~ & ~ & ~ & ~ & ~ & ~ & ~  & \\
 & $K_S K^+$ & 1.450 & 2.473 & 1.80$\pm$0.55 &  
& $K_S K^{*+}$ & 0.704  & 1.096  & 2.15$\pm$0.70 \\
\hline
\end{tabular}
\end{table}

We comment at first on the PP decays. A comparison of the color--favored 
decays (into $\pi^+ \eta$, $\pi^+ K^0$ and $K^+ K^0$) shows that the 
ratios of Cabibbo--forbidden and doubly--Cabibbo--forbidden to the 
Cabibbo--allowed BR are, respectively, about 0.1 and 0.001, instead 
of the expected values ($\tan^2 \theta_C \simeq 0.05$ and 
$\tan^4 \theta_C \simeq 0.003$): this may be explained as the effect of 
a rather large annihilation 
contribution in the Cabibbo--forbidden decay, $D^+_s \to \pi^+ K^0$, and 
of an important interference in the doubly--Cabibbo--forbidden 
channel. Similar arguments hold for the color--suppressed decays. A
discussion of the final state $K^+ \eta$ would be more involved,
since also the non-strange 
components of the $\eta$ meson participate in this decay. 

In the decays to PV final states, as it was noted for the $D^+$, 
those induced by the $VV$ part of the effective Hamiltonian are larger 
than the others in the Cabibbo--allowed and doubly--forbidden cases. 
The Cabibbo--forbidden and color--favored decays $D^+_s \to \rho^+ 
K^0$ and $D^+_s \to \pi^+ K^{*0}$ are roughly equally frequent, as a  
consequence of large annihilation contributions, equal and 
opposite\footnote{The opposite sign arises from the F--type coupling
of the pseudoscalar density.} 
in these two channels. The annihilation terms are also responsible of 
the big difference among the Cabibbo--forbidden and color--suppressed decays
$D^+_s \to$ $\rho^0 K^+$, $\omega K^+$ and $\pi^0 K^{*+}$.
The rescattering corrections are often quite large, and the pattern 
discussed above is somehow hidden in the final results. However, it 
must be said that the 
general quality of the fit for $D^+_s$ decays is not very good,  
for example the prediction for $D^+_s \to \rho^+ \eta'$ is 
${\rm BR_{NR(RR)}}=2.52\;(2.46)\;\%$ against an experimental value 12$\pm$4 \%. 
It may well be that some surprise will come out from new $D^+_s$ data, once 
better statistical and systematical accuracy is attained. 

\begin{table}[ht]
\caption{
Branching ratios (in \%) for some nonleptonic
decays of $D^0$ mesons.}
\vspace{0.3 cm}
\begin {tabular} {| c l c c c | c l c c c |}
\hline
 & $D^0 \to$ & ${\rm BR_{NR}}$ & ${\rm BR_{RR}}$ &  ${\rm BR_{exp}}$ &  
& $D^0  \to$ & ${\rm BR_{NR}}$ & ${\rm BR_{RR}}$ & ${\rm BR_{exp}}$ \\
\hline
 & $\pi^+  K^-$ & 5.114 & 3.847 &  3.80$\pm$0.09 & 
& $\rho^+  K^-$ & 17.029 & 11.201 & 10.2$\pm$0.8 \\
 & ~ & ~ & ~ & ~ &  
& $\pi^+ K^{*-}$ & 2.568 & 4.656 & 6.0$\pm$0.5 \\
 & ${\overline K}^0 \pi^0$ & 0.711 & 1.310 & ---  & 
& ${\overline K}^{*0} \pi^0$ & 1.024 & 3.208 & 2.8$\pm$0.4 \\
 & ~ & ~ & ~ & ~  & 
& ${\overline K}^0 \rho^0$ & 1.607  &  0.759  & --- \\
 & ~ & ~ & ~ & ~ &  
& ${\overline K}^0 \omega$ & 0.345 & 1.855 & --- \\
 & $K^+ K^-$ & 0.579 &   0.424  & 0.412$\pm$0.014 &  
& $K^{*+} K^-$ &  0.839  &   0.431  & 0.38$\pm$0.08 \\
 & ~ & ~ & ~ & ~ &  
& $K^+K^{*-}$ & 0.101  &  0.290  & 0.20$\pm$0.11 \\
 & ~ & ~ & ~ & ~ &  
& $\ \phi\ \pi^0$ & 0.098 & 0.105 & $< 0.14$ \\
 & $K^0 {\overline K}^0$ & 0.0 & 0.130 & 0.071$\pm$0.019 &  
& ${\overline K}^{*0} K^0$ & 0.008 & 0.052 & $< 0.17$ \\
 & ~ & ~ & ~ & ~ &  
& ${\overline K}^0 K^{*0}$ & 0.008 & 0.062 & $< 0.09$ \\
 & $\pi^+ \pi^-$ & 0.500  & 0.151 & 0.143$\pm$0.007 & 
& $\rho^+ \pi^-$ & 1.048  & 0.706  & --- \\
 & ~ & ~ & ~ & ~ &  
& $\pi^+ \rho^-$ & 0.350 & 0.485 & --- \\
 & $\pi^0 \pi^0$ &  0.052  & 0.115 & 0.084$\pm$0.022 & 
& $\rho^0 \pi^0$ & 0.137 & 0.216  & --- \\
 & ~ & ~ & ~ & ~ &  
& $\omega \pi^0$ & 0.010  & 0.013 & --- \\
 & $K^+  \pi^-$ & 0.048  &  0.033  &  0.015$\pm$0.02 & 
& $K^{*+}  \pi^-$ & 0.059  &  0.039  & --- \\
 & ~ & ~ & ~ & ~ &  
& $K^+ \rho^-$  & 0.015 & 0.025 & --- \\
 & $K^0 \pi^0$  &  0.007 &  0.008 & --- & 
& $K^{*0} \pi^0$ & 0.008 &  0.004 &  --- \\
 & ~ & ~ & ~ & ~ &  
& $K^0 \rho^0$ & 0.001 &  0.008 & --- \\
 & ~ & ~ & ~ & ~ &  
& $K^0 \omega$ & 0.004 & 0.002 & --- \\
 & ~ & ~ & ~ & ~ & ~ & ~ & ~ & ~ &  \\
 & $K_S \pi^0$ & 0.428 & 0.759 & 1.14$\pm$0.11 &  
& $K_S \rho^0$ & 0.842  & 0.446  & 0.735$\pm$0.145 \\
 & ~ & ~ & ~ & ~  &  
& $K_S \omega$ & 0.212 & 0.973 & 1.1$\pm$0.2 \\
\hline
\end{tabular}
\end{table}

In Table~3 we have reported for completeness the data and predictions for 
some exclusive $D^0$ decays. Comments analogous to those given on the 
other Tables could be made, in particular concerning the large differences 
between $K^{*+} K^-$ and $K^+ K^{*-}$, as well as between $\rho^+ \pi^-$ 
and $\pi^+ \rho^-$, which are rather softened by rescattering 
effects. We note moreover that final state interactions 
are essential to allow the decays into a pair of strange neutral 
mesons \cite{Lipkin:es,Pham:1987rj}. 

It may be of some interest to consider partially inclusive rates of 
Cabibbo--allowed (CA), forbidden (CF) and doubly--forbidden (DCF) decay 
processes.
For such quantities the comparison of experimental data with 
theoretical predictions are much less dependent on the rescattering 
corrections. Their ratios may then be compared with the appropriate 
power of $\tan \theta_C$ to check if the simple--minded suppression 
(CF/CA $\simeq 2\;\tan^2 \theta_C$ and DCF/CF $\simeq 0.5\;\tan^2 
\theta_C$) gives an adequate description of the data. 
This is illustrated in Table~4, that we now comment upon.

The success of the theoretical model in describing the data is 
generally acceptable. The worst case is given by the Cabibbo--allowed 
$D^+ \to {\rm PV}$ decays, where the theoretical predictions are considerably 
larger than experiment. 

Concerning the  
inclusive data and theoretical predictions for $D^0$ decays, the 
simple--minded suppression referred to above is a consequence of 
(flavor) SU(3) symmetry. In the limit of total decoupling of the third 
quark family, 
the weak effective hamiltonians form a U-spin triplet, the $D^0$ 
meson has zero U-spin and therefore one has only one decay amplitude 
\cite{Kingsley:fe,Voloshin:1975yx,
Buccella:1994nf,Lipkin:2000sf,Gronau:2000ru}. Moreover, the annihilation 
contributions to 
decays into PP vanish for CF, are SU(3) violating and of opposite 
sign for CA and DCF, while they are nonvanishing and SU(3) allowed 
for decays to PV. 
These features give a reasonable explanation of the results 
for the ratios, both bigger than naively expected in PP decays and 
near to the naive expectations in PV decays. 

\begin{table}[t]
\caption{
Inclusive branching ratios (in \%) for two-body 
decays of $D$ mesons.}
\vspace{0.3 cm}
\begin {tabular} {| c l c | c c c c | c c c |}
\hline
  &  & &  & CA  & CF & DCF & &  CF/CA & DCF/CF \\
\hline
 & $D^0 \to {\rm PP}$ & exp. & &  8.71$\pm$0.38 & $> 0.68$ & & &  $> 0.075$ & \\ 
 & &  th.  &  &  7.34 & 1.31 & 0.057 & &  0.18 & 0.043 \\
  &  &  &  & & & & & & \\
 & $D^0 \to {\rm PV}$ & exp. & &  25.4$\pm$1.2 &  $> 0.44$ & & &  $>0.02$ & \\
  & & th. & &  23.0 & 2.68 & 0.085 & &  0.12 & 0.032 \\
\hline
 & $D^+ \to {\rm PP}$ & exp. &  & 2.77$\pm$0.18 & $> 1.43$ &  &  & $> 0.44$ &  \\
  &  & exp. & & & $< 2.56$ &  & &$< 0.92$ & \\
  & & th. & &  2.94 & 2.10 & 0.164 &  & 0.71 & 0.078 \\
  &  &  & & & & & & & \\
 & $D^+ \to {\rm PV}$ & exp.  & & 8.5$\pm$2.5 & $> 2.6$ &   & & $> 0.24$ & \\
  &   & th. & &  14.2 & 3.35 & 0.205  & & 0.24 & 0.061 \\
\hline
 & $D^+_s \to {\rm PP}$ & exp. &  & 9.2$\pm$1.6 & & &  & & \\
  & & th.  & & 11.2 & 1.31 & 0.012 &  & 0.12 & 0.009 \\
  &  &  &  & & & & & & \\
 & $D^+_s \to {\rm PV}$ & exp. &  & 32.4$\pm$4.6 &  &  &  &  &  \\
  & & th. & &  25.1 & 2.37 & 0.024  & & 0.094 & 0.010 \\
\hline
\end{tabular}
\end{table}

The charged and charmed mesons form a U-spin doublet. As a 
consequence, even in the symmetric limit, we have two independent decay 
amplitudes, and one would recover the simple-minded suppression 
factors only if these two amplitudes were equally important. More 
generally, SU(3) symmetry would predict the following relations
\begin{equation}
\left(\frac{{\rm CF}}{{\rm CA}}\right)_{D^+} \cdot \;
\left(\frac{{\rm DCF}}{{\rm CF}}\right)_{D^+_s}\;=\;
\left(\frac{{\rm CF}}{{\rm CA}}\right)_{D^+_s} \cdot \;
\left(\frac{{\rm DCF}}{{\rm CF}}\right)_{D^+} \;=\; \tan^4 \theta_C\;, 
\end{equation}
both for PP and for PV final states. From Table~4 one 
obtains for the three terms in the above equation the numerical results 
(0.0064, 0.0094, $\tan^4 \theta_C=0.0026$) for PP decays and 
(0.0024, 0.0057, 0.0026) for PV decays. This shows that 
the flavour symmetry is rather badly broken in this factorized model, 
without any need of ``new physics''. 

The inclusive data of $D^+$ show once more that 
the ratio CF/CA is much larger than the (very) naive expectation, due to 
destructive interference in the Cabibbo--allowed amplitudes. 
This is particularly evident for the decays into PP, although no 
single branching fraction is larger than 1\%.  
The ratio DCF/CF is also larger than expected, due to the annihilation 
contribution in DCF. 

The predictions for $D^+_s$ inclusive decays (the data are 
still too sparse to give indications) suggest that, while CF/CA is 
``normal'', the ratio DCF/CF is smaller than naively expected, due to 
interference effects in DCF.       

Decays in two vector mesons are not 
included in this fit, so that we are not able to
present a theoretical estimate of  
the other large BR($D^+ \to K^{*+}{\overline K}^{*0}$). 

In this paper, we have shown that in a model for $D$ decays based on 
(generalized) factorization and including resonance--mediated final 
state interactions it is possible to get a branching 
ratio for the Cabibbo--forbidden channel $D^+ \to K^{*+}{\overline K}^0$ 
of about 1.52 \% versus an experimental value of 
3.1$\pm$1.4 \% (in fact, an even larger prediction for this BR 
was given in a previous version of the model \cite{Buccella:1994nf} 
prior to the experimental result). On the other hand, the theoretical 
prediction for the Cabibbo--allowed decay BR($D^+ \to \rho^+ K_S$) 
is  5.82 \%  versus the experimental result  
3.30$\pm$1.25 \% (the other Cabibbo--allowed decay is well reproduced). 
Within the present, large error bands this is compatible: for the 
ratio of the two BR's we have 0.26 (theory) versus 0.94$\pm$0.55 
(experiment). While the central values undeniably disagree, the discrepancy 
may be alleviated not only by a much smaller BR($D^+ \to K^{*+}{\overline K}^0$), 
but also by a somewhat larger BR($D^+ \to \rho^+ K_S$): this is the 
direction to which a model using factorization and no extreme ``new 
physics'' hypothesis would point.

\end {document}